\documentclass[a4paper,onecolumn,11pt]{quantumarticle}
\pdfoutput=1

\usepackage[utf8]{inputenc}
\usepackage{lipsum}

\usepackage{amsmath, amssymb,amsfonts,amsthm,mathtools}
\usepackage{xspace,graphicx,relsize,bm,xcolor}
\usepackage{soul} 

\usepackage{bbold}
\usepackage{xcolor}
\usepackage{multicol}
\usepackage{csquotes}

\usepackage{parskip}  

\usepackage{dsfont}
\usepackage[T1]{fontenc}
\usepackage{enumitem}

\usepackage[
    backend=biber,
    style=numeric-comp, 
    sorting=anyt,
    minalphanames=3,
    maxalphanames=3,
    maxnames=99,
    backref=true
    ]{biblatex}
\DefineBibliographyStrings{english}{%
  backrefpage = {page},
  backrefpages = {pages},
}

\usepackage{sepfootnotes}
\newendnotes{x}
\renewcommand\xnotesize\normalsize

\usepackage{hyperref}
\usepackage[margin=3.0cm]{geometry}
\definecolor{linkcol}{rgb}{0.0,0.55,0.7}
\definecolor{citecol}{rgb}{0.0, 0.6, 0.45}
\definecolor{urlcol}{rgb}{0.7, 0.0, 0.55}
\hypersetup{
	colorlinks,
	linkcolor={linkcol},
	citecolor={citecol},
	urlcolor={urlcol}
}

\usepackage{url}
\usepackage{subcaption}
\usepackage{mleftright}
\usepackage{hyperref}
\usepackage{multirow}
\usepackage{physics}  

\usepackage{algorithm}
\usepackage{algpseudocode}

\usepackage{cleveref}

\usepackage{authblk}  

\newcommand{\probP}{\text{I\kern-0.15em P}}

\providecommand{\myvec}[1]{\ensuremath{\boldsymbol{#1}}}

\providecommand{\xx}{\ensuremath{\myvec{x}}}
\providecommand{\yy}{\ensuremath{\myvec{y}}}
\providecommand{\zz}{\ensuremath{\myvec{z}}}

\providecommand{\ttheta}{\ensuremath{\myvec{\theta}}}

\providecommand{\bbE}{\ensuremath{\mathbb{E}}}

\newtheoremstyle{mydefinitionsty}
{10pt}
{10pt}
{}
{}
{}
{}
{.5em}
{\textbf{\thmname{#1}~\thmnumber{#2}:  }\thmnote{(#3)}}
\theoremstyle{mydefinitionsty}
\newtheorem{definition}{Definition}

\newtheoremstyle{myproblemsty}
{10pt}
{10pt}
{}
{}
{}
{}
{.5em}
{\textbf{\thmname{#1}~\thmnumber{#2}:  }\thmnote{(#3)}\newline}
\theoremstyle{myproblemsty}

\newtheoremstyle{mythmsty}
{10pt}
{10pt}
{\itshape}
{}
{}
{}
{.5em}
{\textbf{\thmname{#1}~\thmnumber{#2}:  }\thmnote{(#3)}}
\theoremstyle{mythmsty}

\newtheorem{proposition}{Proposition}

\newcommand{\python}[1]{\texttt{#1}}

\title{IQPopt: Fast optimization of instantaneous quantum polynomial circuits in JAX}

\author[1,2]{Erik Armengol\thanks{erik.recio@icfo.eu}}
\author[3]{Joseph Bowles\thanks{joseph@xanadu.ai}}

\affil[1]{ICFO-Institut de Ciencies Fotoniques, The Barcelona Institute of Science and Technology, 08860, Castelldefels (Barcelona), Spain}
\affil[2]{Eurecat, Centre Tecnològic de Catalunya, Multimedia Technologies, 08005 Barcelona, Spain}
\affil[3]{Xanadu, Toronto, ON, M5G 2C8, Canada}

\date{7 May 2026}

\begin{document}

\makeatletter

\maketitle


\begin{abstract}
    IQPopt is a software package designed to optimize large-scale instantaneous quantum polynomial circuits on classical hardware. By exploiting an efficient classical simulation algorithm for expectation value estimation, circuits with thousands of qubits and millions of gates can be optimized, provided the relevant objective function has an efficient description in terms of Pauli-Z type observables. Since sampling from instantaneous quantum polynomial circuits is widely believed to be hard for classical computers, this provides a method to identify powerful circuit instances before deployment and sampling on quantum hardware, where computational advantages may exist. The package leverages automatic differentiation in JAX, can be accelerated with access to hardware accelerators such as graphics processing units, and contains a dedicated module that can be used to train and evaluate quantum generative models via the maximum mean discrepancy. 
\end{abstract}

\begin{figure*}[t]
    \centering
    \includegraphics[width=1.0\linewidth]{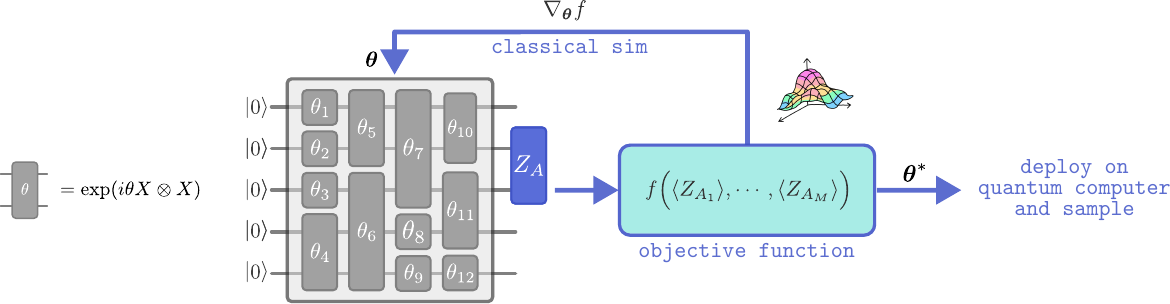}
    \caption{Overview of IQP circuit optimization in IQPopt (\href{https://github.com/XanaduAI/iqpopt}{github.com/XanaduAI/iqpopt}). A parameterized IQP circuit consists of parameterized gates $\exp(i\theta_j X_j)$, with $X_j$ a tensor product of Pauli X operators. The objective function must be expressed as a differentiable function of expectation values of products of Pauli-Z operators evaluated at the output of the circuit. By exploiting a fast classical algorithm for these expectation values, circuit parameters can be optimized in IQPopt for circuits with thousands of qubits. After optimization, the optimal circuit can then be deployed on a quantum computer, where sampling can lead to computational advantages relative to fully classical algorithms.}
    \label{fig:iqpopt}
\end{figure*}

\section{Introduction}
Instantaneous Quantum Polynomial (IQP) circuits are a commonly studied class of quantum circuits comprised of commuting gates, and have become well known due to a number of results related to the hardness of sampling from their output distributions \cite{iqp_mult, iqp_add, iqp_add2}. In particular, the strongest of these results was recently proven in \cite{iqp_add2} (a strengthening of \cite{iqp_add}), showing that the ability to classically sample from such circuits to additive error would imply the collapse of the polynomial hierarchy to its second level (subject to one of two additional conjectures). As a  result, IQP circuits are prime candidates for potential quantum advantages for tasks related to sampling, and have been the focus of a number of theoretical \cite{iqp_exp1,iqp_exp2,iqp_exp3} and experimental \cite{iqp_exp4,iqp_lukin} works aiming towards this goal.

\begin{figure*}
    \centering
    \includegraphics[width=1.0\textwidth]{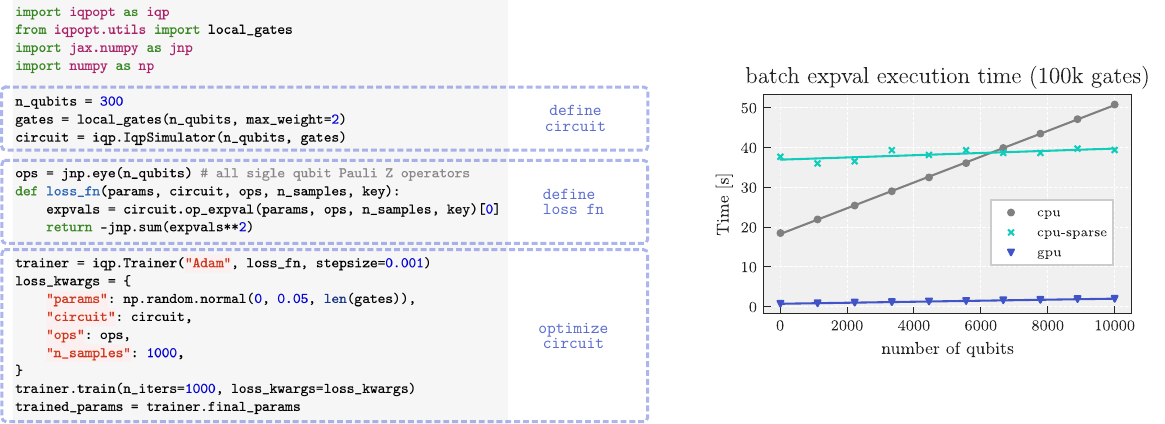}
    \caption{(left): An example code block that defines a 300 qubit circuit with 45150 parameters and optimizes a simple loss function. (right) Time taken to estimate 10000 randomly chosen expectation values (up to additive error $\approx$ 0.01) for a circuit with 100000 gates as a function of the number of qubits in the circuit. Calculations were performed on a Nvidia Grace Hopper GH200 chip, using either CPU hardware, CPU hardware and sparse matrix operations, or GPU hardware. Further benchmarks are shown in Sec.~\ref{sec:benchmarks}.}
    \label{fig:intro}
\end{figure*}

To assess the performance of a circuit at a sampling task, one generally needs to sample from the output distribution of the circuit. This is the case, for example, for the linear cross-entropy benchmark \cite{xeb} used to evaluate IQP circuit sampling in \cite{iqp_lukin}. In certain cases however, the performance at a sampling task can be inferred from correlations alone, i.e., from the expectation values of products of Pauli Z operators. This situation arises in combinatorial quantum optimization algorithms \cite{qaoa, qaoa2, qaoa3}, where a diagonal Ising Hamiltonian measures the average objective function value of the solutions obtained from sampling the circuit. Another use case arises in generative quantum machine learning,  where the quality of samples as assessed by the maximum mean discrepancy can be efficiently estimated via correlations of subsets of bits \cite{rudolph2024trainability}. In these cases, one does not need to sample from the circuit to assess performance, so long as other methods to estimate expectation values are available. Interestingly, such a situation occurs for the class of IQP circuits. In particular, these circuits admit a curious form of computational complexity: whereas sampling is expected to be hard, efficient classical algorithms are known for the task of estimating expectation values \cite{nest2010simulating} up to inverse polynomial error\footnote{Inverse polynomial error estimates of probabilities can also be obtained following \cite{shepherd2010binary, pashayan2020estimation}, which is exploited in \cite{kyriienko2024protocols} in a similar fashion to this work.}. Consequently, for problems with the aforementioned structure, the performance of IQP circuits can be assessed with classical hardware alone, although this is a relatively unexplored area.

In this work we present a software package, called IQPopt, that exploits this property in order to optimize large-scale IQP circuits. In particular, the package is designed to optimize parameterized classes of IQP circuits with respect to objective functions that can be expressed as functions of Pauli-Z type expectation values (see Fig.~\ref{fig:iqpopt}). After optimization, the optimal parameters can then be used to deploy the circuit on quantum hardware and generate samples, potentially leading to advantages relative to purely classical algorithms. We implement a version of a simulation algorithm contained in \cite{nest2010simulating} in JAX \cite{jax} to arrive at an optimization algorithm that uses automatic differentiation to compute gradients, which are then used in standard gradient descent methods to minimize the objective function (see Fig.~\ref{fig:intro} for example code). This implementation allows for the possibility of optimizing circuits with thousands of qubits and millions of gates with relatively modest hardware requirements. Moreover, the optimization can be accelerated significantly with GPU access (see Fig.~\ref{fig:intro}, right). For tasks that fit this framework, this therefore represents an enormous reduction in the effective cost and time needed to optimize these circuits, and opens the door to empirical studies on the power of large scale IQP circuits. The package also allows one to optimize a class of classical circuits that correspond to replacing the gates of a parameterized IQP circuit by their de-cohered classical analogues. These circuits can therefore be used as a classical performance comparison in order to provide evidence that interference is being used non-trivially in the circuit. 

The development of the package was motivated by an empirical study that investigates the prospect of using IQP circuits as generative quantum machine learning models, which we present in an upcoming work \cite{upcoming}. For this reason, aside from the core optimization functionality, the package also contains a module that is specifically targeted at generative quantum machine learning, and provides tools to train parameterized IQP circuits via the maximum mean discrepancy, and evaluate their performance using the kernel generalized empirical likelihood \cite{ravuri2023understandingdeepgenerativemodels}. We hope this can be a valuable tool for the quantum machine learning community, and that the package finds applications in other areas of quantum computation.

The rest of the article is structured as follows. In section \ref{sec:theory} we cover the theoretical preliminaries underpinning the code as well as its time and space complexity. In section \ref{sec:overview} we explain how to use the package with example code, and in section \ref{sec:benchmarks} we perform a number of benchmarks to assess typical runtimes.

\section{Theoretical preliminaries}\label{sec:theory}
In this section we cover the theoretical aspects that are necessary to understand the inner workings of the code, as well as an analysis of runtime and memory scaling. We start by defining the class of circuits and optimization problems to which the package applies. 

\begin{definition}[parameterized IQP circuit]
An $n$-qubit \textit{parameterized IQP circuit} is a circuit comprised of 
\begin{enumerate}
    \renewcommand{\labelenumi}{(\roman{enumi})}
    \item State initialisation in $\ket{0}$.
    \item Parameterized gates of the form $\exp(i\theta_j X_{\myvec{g}_j})$, where $X_{\myvec{g}_j}$ is a tensor product of Pauli X operators acting on a subset of qubits specified by the non-zero elements of $\myvec{g}_j\in\{0,1\}^n$.
    \item Measurement in the computational basis.
\end{enumerate}
\end{definition}
The full parameterized unitary is thus $U(\ttheta)=\prod_j \exp(i\theta_j X_{\myvec{g}_j})$ where $\ttheta = (\theta_1, \cdots, \theta_m)$ denotes the vector of optimization parameters. Note that since the generators commute, the order in which the gates are applied is irrelevant. 

IQPopt is designed to minimize objective functions that can be expressed as differentiable functions of Pauli Z type expectation values. Similarly to $X_{\myvec{g}_j}$, we denote by $Z_{\myvec{a}_j}$ a tensor product of Pauli Z operators where the non-zero elements of $\myvec{a}_j\in\{0,1\}^n$ denote on which qubits a Pauli Z operator acts. The relevant problems then take the following form.
\begin{definition}[IQPopt problem instance]\label{def:iqpopt}
    Given a parameterized IQP circuit with parameters $\ttheta$, a collection of observables $\{ \langle Z_{\myvec{a}_1} \rangle , \cdots ,\langle Z_{\myvec{a}_l}\rangle \}$ and a continuous and differentiable function $f$,  
    \begin{align}\label{eqn:opt}
    \underset{\ttheta}{\text{minimize }}  f\left(\langle Z_{\myvec{a}_1} \rangle , \cdots ,\langle Z_{\myvec{a}_l}\rangle\right),
    \end{align}
where
\begin{align}\label{eqn:expvals}
    \langle Z_{\myvec{a}_j} \rangle &= \bra{0}U^\dagger(\ttheta)Z_{\myvec{a}_j} U(\ttheta)\ket{0}
\end{align}
    is the expectation value of $Z_{\myvec{a}_j}$ evaluated at the output of the circuit. 
\end{definition}

Note that since expectation values provide an alternative basis to probabilities to describe the output distribution of the circuit (\cite{o2021analysis} see Theorem 1.1), any differentiable function of the circuit output probabilities can be expressed in the above form at the cost of exponentially many observables. The problems best suited to IQPopt are therefore those that admit an efficient description in terms of correlations rather than probabilities.

\subsection{Efficient estimation of expectation values}
Here we show how to efficiently estimate expectation values of the form \eqref{eqn:expvals} on classical hardware, to inverse polynomial additive error. This is the same precision achieved by sampling the quantum circuit, in the sense that sampling from $U(\ttheta)\ket{0}$ and estimating expectation values from the resulting statistics also results in an inverse polynomial additive error with respect to the number of samples. A classical algorithm that achieves this is contained in the results of \cite{nest2010simulating}, which we use here in a form that applies directly to our circuits. 

\begin{proposition}\label{prop:expval}
    (follows from \cite{nest2010simulating}, Theorem 3) Given a parameterized IQP circuit on $n$ qubits, an expectation value $\langle Z_{\myvec{a}}\rangle$ and an error $\epsilon=\text{poly}(n^{-1})$, there exists a classical algorithm that requires $\mathcal{O}(\text{poly}(n))$ time and space, and samples a random variable that is an unbiased estimator of $\langle Z_{\myvec{a}}\rangle$, with standard deviation less than $\epsilon$. 
\end{proposition}

\noindent\textit{Proof}: Inserting identities $\mathbb{I}=H^2$ (with $H$ an $n$-fold tensor product of Hadamard unitaries) between operators in \eqref{eqn:expvals} the expression becomes
\begin{align}
    &\bra{0}H (H U^{\dagger}(\ttheta) H)(H Z_{\myvec{a}} H)(H U(\ttheta)H)H\ket{0} \nonumber \\ &\quad\quad= 
    \bra{+}^{\otimes n} D^{\dagger}(\ttheta) X_{\myvec{a}} D(\ttheta)\ket{+}^{\otimes n}, \label{eq:bowles_hadamard}
\end{align}
with $D(\ttheta)$ a diagonal unitary analogous to $U(\ttheta)$ where the generators are now Pauli Z tensors,
\begin{align}\label{eq:diagcircuit}
    D(\ttheta)=\prod_{j}e^{i\theta_j Z_{\myvec{g}_j}}.
\end{align}
Since this operator is diagonal, the eigenvectors of $D(\ttheta)$ are computational basis states. The corresponding eigenvalues $\lambda_{\zz}$ are easy to compute since a basis state $\ket{\zz}$ picks up a phase $\exp(\theta_{j}(-1)^{\myvec{g}_{j}\cdot\zz})$ for each gate in \eqref{eq:diagcircuit}:
\begin{align}\label{eq:eigenvals}
    D(\ttheta)\ket{\zz}=\exp(i\sum_{j} \theta_{j}(-1)^{\myvec{g}_{j}\cdot\zz})\ket{\zz} = \lambda_{\zz}\ket{z}.
\end{align}
Writing $\ket{+}^{\otimes n}=2^{-n/2}\sum_{\zz}\ket{\zz}$ in \eqref{eq:bowles_hadamard} and using the above and $X_{\myvec{a}}\ket{\zz}=\ket{\zz\oplus {\myvec{a}}}$ we arrive at 
\begin{align}
     \langle Z_{\myvec{a}} \rangle = \frac{1}{2^n} \sum_{\zz} \lambda^{*}_{\zz\oplus{\myvec{a}}}\lambda_{\zz} 
    =\frac{1}{2^n} \sum_{\zz}\Re\left[\lambda^{*}_{\zz\oplus\myvec{a}}\lambda_{\zz}\right], \label{eq:zaq}
\end{align}
where we have taken the real component since $\langle Z_{\myvec{a}} \rangle$ is guaranteed to be real. Expanding the expression via \eqref{eq:eigenvals}  and using the fact that \eqref{eq:zaq} is an expectation value over bit strings we then find 

\begin{align}
    \langle Z_{\myvec{a}} \rangle &=\frac{1}{2^n}\sum_{\zz} \Re\left[e^{-i\sum_{j=1}^{m}\theta_j (-1)^{\myvec{g_j}\cdot \zz\oplus\myvec{a}}}e^{i\sum_{j=1}^{m}\theta_j (-1)^{\myvec{g_j}\cdot\zz}} \right]\\
    &=\mathbb{E}_{\zz\sim U}\Big[ \cos \Big(\sum_{j=1}^{m}\theta_j \left((-1)^{\myvec{g_j}\cdot\zz} - (-1)^{\myvec{g_j}\cdot \zz\oplus\myvec{a}}\Big)\right)\Big]\\
    &=\mathbb{E}_{\zz\sim U}\Big[ \cos\Big(\sum_j \theta_{j}(-1)^{\myvec{g}_{j}\cdot \zz}(1-(-1)^{\myvec{g}_j\cdot \myvec{a}}\Big) \Big].
    \label{eqn:op_expval}
\end{align}
This allows us to compute unbiased estimates of $\langle Z_{\myvec{a}} \rangle$ efficiently by replacing the expectation with an empirical mean. That is, if we sample a batch of $s$ bit strings $\{\zz_i\}$ from the uniform distribution and compute the sample mean
\begin{align}\label{eqn:expvalsample}
    \hat{\langle Z_{\myvec{a}}\rangle} = \frac{1}{s}\sum_{i}\cos\Big(\sum_j \theta_j(-1)^{\myvec{g}_j\cdot \zz_i}(1-(-1)^{\myvec{g}_j\cdot \myvec{a}})\Big),
\end{align}
we obtain an unbiased estimate $\hat{\langle Z_{\myvec{a}}\rangle}$ of $\langle Z_{\myvec{a}}\rangle$; i.e.\ we have that $\mathbb{E}[\hat{\langle Z_{\myvec{a}}\rangle}] = \langle Z_{\myvec{a}}\rangle$. Since by the central limit theorem, the variance of the sample mean of a bounded random variable decreases as $1/s$, it follows that by taking $s=\mathcal{O}(\text{poly}(n))$ we obtain an estimator for $\langle Z_{\myvec{a}} \rangle$ with standard deviation less than $\epsilon=\text{poly}(n^{-1})$. $ \blacksquare$

\subsection{Efficient batch evaluation of expectation values}\label{sec:batchexpval}
Equation \eqref{eqn:expvalsample} shows us how to get efficient estimates of single expectation values $\langle Z_{\myvec{a}} \rangle$ through simple trigonometric and linear algebra operations. Here we show how to exploit this structure to obtain a fast algorithm to batch evaluate large numbers of expectation values via matrix multiplications. We first define the following matrices: 
\begin{align}
    G &= (\myvec{g}_1, ..., \myvec{g}_m)^T \rightarrow m \times n \text{ matrix,}\\
    Z &= (\zz_1, ..., \zz_s)^T \rightarrow s \times n \text{ matrix,}\\
    A &= (\myvec{a}_1, ..., \myvec{a}_l)^T \rightarrow l \times n \text{ matrix,}\\
    \Theta &= \text{diag}(\theta_1, ..., \theta_m) \rightarrow m \times m \text{ matrix},
\end{align}
with $\{\myvec{a}_1,\cdots,\myvec{a}_l\}$ describing the batch of expectation values we wish to estimate and $\{\zz_1,\cdots\zz_s\}$ a set of $s$ bit strings as in \eqref{eqn:expvalsample}. We now write \eqref{eqn:expvalsample} as 
\begin{align}
\hat{\langle Z_{\myvec{a}_k}\rangle} &= \frac{1}{s}\sum_{i}\cos\Big(\sum_j \theta_j(-1)^{\myvec{g}_j\cdot \zz_i}(1-(-1)^{\myvec{g}_j\cdot \myvec{a}_k})\Big)\\
&= \frac{1}{s}\sum_{i}\cos\Big(\sum_j (-1)^{ \zz_i\cdot\myvec{g}_j} (1-(-1)^{ \myvec{a}_k\cdot\myvec{g}_j})\theta_j\Big)\\
&= \frac{1}{s} \sum_{i}\cos\Big(\sum_j B_{i,j} \cdot C_{k,j}\,\theta_j\Big)\\
&= \frac{1}{s} \sum_{i}\cos\Big( E_{k,i}\Big). \label{eqn:batchalgo}
\end{align}
where we have defined the matrices $B$, $C$ and $E$
\begin{align}
    B_{i,j} &= (-1)^{\zz_i\cdot\myvec{g}_j}\iff B = (-1)^{Z \cdot G^T} \rightarrow s \times m \text{ matrix,}\\
    C_{k,j} &= (1-(-1)^{\myvec{a}_k\cdot\myvec{g}_j}) \iff C = (1 - (-1)^{A \cdot G^T}) \rightarrow l \times m \text{ matrix,}\\
    E_{k,i} &= \sum_j C_{k,j}\cdot \theta_j \cdot B_{i,j} \iff E = C \cdot \Theta \cdot B^T \rightarrow l \times s \text{ matrix.} \label{eqn:Emat}
\end{align}
From this we can see that the batch estimation of expectation values amounts to calculating the matrix $E$, applying an element wise cosine function and taking column-wise means. The most computationally intensive parts of this calculation are the matrix multiplications $Z \cdot G^T$, $A \cdot G^T$ and $C \cdot \Theta \cdot B^T$, resulting in a runtime $\mathcal{O}(m\cdot(s\cdot n + l\cdot n + s\cdot l))$ using standard matrix multiplication algorithms. Putting this all together and using the fact that the standard error in \eqref{eqn:expvalsample} scales as $1/\sqrt{s}$ we have the following proposition. 
\begin{proposition}[Batch evaluation of expectation values]\label{prop:expval_batch}
Given a parameterized IQP circuit on $n$ qubits with $m$ parameters, and a list of $l$ observables $\{Z_{\myvec{a}_1}, \cdots,  Z_{\myvec{a}_l}\}$, there exists a classical algorithm that returns unbiased estimates of the expectation values $\{\langle Z_{\myvec{a}_1}\rangle, \cdots,  \langle Z_{\myvec{a}_l}\rangle\}$, each with standard deviation less than $\epsilon$, in $\mathcal{O}(m\cdot(\frac{n+l}{\epsilon^2} + l\cdot n))$ runtime.  
    
\end{proposition}

An important point to notice is that we have used the same batch of bitstrings $Z$ in the calculation of each $\hat{\langle Z_{\myvec{a}_k}\rangle}$. Following \eqref{eqn:expvalsample} each estimator will be unbiased, but the use of the same $Z$ will generally result in correlations between the estimates which may need to be taken into account. We note that a similar effect also happens when using samples from a quantum circuit to batch estimate expectation values, since the estimates are functions of a set of common samples from the circuit. Uncorrelated estimates can of course be obtained by using a tensor of shape $l\times s \times n$ in place of $Z$, at the cost of a factor $l$ in runtime. Such an option is available in the package, as we describe in Sec.~\ref{sec:expvals}.

\subsection{Controlling memory}\label{sec:memory}
The memory cost of batch estimation of expectation values is dictated by the size of the matrices that need to be stored and thus results in a space complexity
\begin{align}
    \mathcal{O}(\max(nm, ln, sn, ms, ml, sl))
\end{align}
With 32GB of RAM, this leads to maximum values around $10^{10}$ for any product of two of $n,m,s,l$. In many applications however, it may be that $G$ or $A$ is sparse. Such a situation arises, for example, if the gate generators or observables are mostly low weight operators so that $\myvec{g}_j$ and $\myvec{a}_j$ are sparse vectors. In such cases, it can be useful to store the generators $G$ and/or the operators $A$ in a sparse representation and use sparse matrix multiplication. The package includes an option for this that uses the \texttt{scipy.sparse} package \cite{2020SciPy-NMeth} and \texttt{csr matrix} format, as we explain in Sec.~\ref{sec:circuit}. As we detail in App.~\ref{app:sparse}, this results in an implementation with space complexity 
\begin{align}
    \mathcal{O}(\max(\text{nnz}(G), \text{nnz}(A), sn, ms, ml, sl)), 
\end{align}
where $\text{nnz}(G)$ is the number of non-zero entries of the sparse matrix $G$. In scenarios where either $nm$ or $ln$ is large, this can therefore be advantageous from a memory perspective. The runtime complexity of the sparse implementation is 
\begin{align}
    \mathcal{O}(l\max\left(\text{nnz}_{\text{row}}(A), \text{nnz}_{\text{row}}(G)\right)^2 + s n^2 + sml),
\end{align}
where $\text{nnz}_{\text{row}}$ denotes the maximum number of non-zero elements in any row. 

Another way to control memory usage is to chunk the batch evaluation of the expectation values $\langle Z_{\myvec{a}}\rangle$ into smaller sub-batches. That is, rather than work with the matrix $A$, chunk $A$ row-wise into sub-batches, compute $E$ for each sub-batch and then concatenate the results. This reduces the effective value of $l$ at the cost of larger runtime. In a similar fashion, note that from \eqref{eqn:expvalsample}, the estimation of each expectation value is a mean with respect to the bit strings $Z$. One can therefore also batch $Z$ into smaller sub-batches, compute $E$ for each and take the mean of the resulting matrices. This reduces the effective value of $s$ at the cost of longer runtime. The package includes options for batching of $A$, $Z$, or both, as explained in Sec.~\ref{sec:expvals}.

\begin{figure*}
    \centering
    \includegraphics[width=0.5\linewidth]{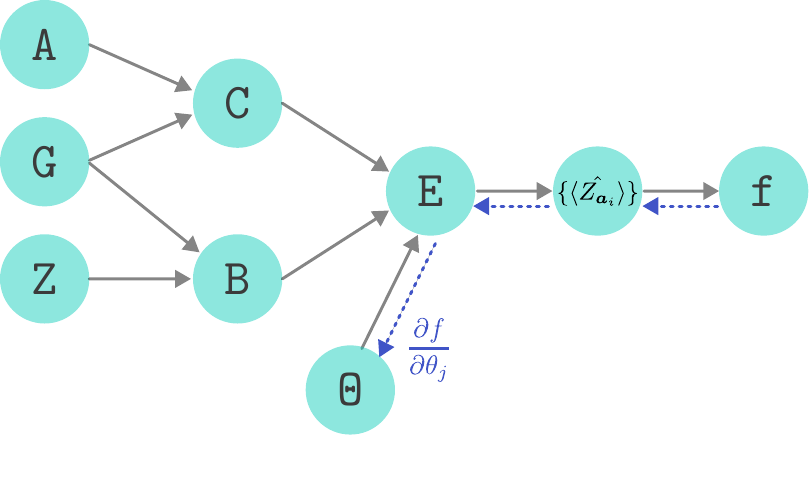}
    \caption{The computational graph computed by IQPopt. The matrices $A$, $G$ and $\Theta$ specify the operators $Z_{\myvec{a}_i}$, generators $X_{\myvec{g}_j}$ and optimization parameters $\ttheta$, and $f$ is the function appearing in the problem instance definition (Def.~\ref{def:iqpopt}). $Z$ contains a batch of randomized bit strings that result in unbiased estimates of the expectation values $\langle Z_{\myvec{a}}\rangle$ via \eqref{eqn:expvalsample} at the second to last node. In each step of optimization, a specific batch of bit strings $Z$ is sampled, and gradients are computed via automatic differentiation by the chain rule (dotted backwards lines).}
    \label{fig:comp_graph}
\end{figure*}

\subsection{Gradient evaluation and optimization in JAX}
We make use of the python package JAX \cite{jax} to perform computations, opting for 64-bit float precision. 
Since JAX excels at linear algebra calculations, it is ideally suited for our purposes. To find solutions to the optimization problem \eqref{eqn:opt} we employ the package JAXopt \cite{jaxopt}, which leverages gradient descent based methods to optimize functions written in JAX. Note that given fixed $G, Z$ and $A$, the batch evaluation algorithm given by \eqref{eqn:batchalgo} is a deterministic and differentiable function of $\Theta$. This means that gradients of the objective function $f\left(\langle Z_{\myvec{a}_1} \rangle , \cdots ,\langle Z_{\myvec{a}_l}\rangle\right)$ can be estimated via automatic differentiation in JAX, at a comparable computational cost to evaluating the objective (see Fig.~\ref{fig:comp_graph} for an image of the computational graph). In particular, at each step of gradient descent, we use a new batch of bit strings $Z$ to estimate the gradient. To perform the optimization, it follows that all that is needed from the user is to code the objective function, as we will see in Sec.~\ref{sec:circuitopt}. 

The package makes use of JAX's just-in-time (JIT) compilation functionality, which uses XLA to optimize and compile Python functions to enable faster execution, which can greatly speed up optimization. Finally, we note that JAX offers seamless integration with GPU and TPU devices. Since the compute intensive operations are matrix multiplications, the code can greatly benefit from access to such devices, as we show in Sec.~\ref{sec:benchmarks}. Utilizing GPU and TPU hardware is straightforward, and only requires that JAX has such a device available and recognized. 

\subsection{Classical stochastic circuit comparisons}\label{sec:bitflip}
One important challenge when benchmarking quantum algorithms is designing meaningful comparisons to classical alternatives \cite{schreiber2023classical, bowles2024better, sweke2023potential, fontana2023classical, basilewitsch2024quantum}. Interestingly, the structure of parameterized IQP circuits lends itself to a natural classical circuit that can be used for such comparisons. The classical circuit in question uses incoherent versions of the parameterized quantum gates and takes the form of a stochastic Markov chain acting on an initial fixed bit string. These classical circuits may therefore be particularly useful in understanding whether coherence is being used by the quantum circuit, by comparing the objective function values of the optimized classical and quantum circuits.

To construct the classical circuit, we first note that the action of the quantum gate $\exp(i \theta_j X_{\myvec{g}_j})$ on a computational state $\ket{\xx}$ is
\begin{align}
    \exp(i \theta_{_j} X_{\myvec{g}_j})\ket{\xx} = \cos(\theta_j)\ket{\xx}+i\sin(\theta_j)\ket{\xx\oplus\myvec{g}_j}.
\end{align}
That is, the gate flips some of the bits of $\xx$ in coherent superposition such that the probability of observing the flipped bit string $\xx\oplus\myvec{g}_j$ from a subsequent measurement is $\sin^2(\theta_j)$. An incoherent version of this gate is therefore one that flips the bits with the same probabilities in a classical stochastic manner:
\begin{align}\label{eq:bitflip}
    \ket{\xx}\bra{\xx} \rightarrow \cos^2(\theta_j)\ket{\xx}\bra{\xx} + \sin^2(\theta_j)\ket{\xx\oplus\myvec{g}_j}\bra{\xx\oplus\myvec{g}_j}.
\end{align}
We will thus call these circuits \emph{stochastic bitflip circuits}. Here we will show how the expectation values $\langle Z_{\myvec{a}}\rangle$ of stochastic bitflip circuits relate to those of paramterized IQP circuits, which will lead us to an efficient algorithm for batch evaluation, in a similar spirit to Sec.~\ref{sec:batchexpval}:

\begin{proposition}
Given a stochastic bitflip circuit on $n$ qubits with $m$ parameters, and a list of $l$ observables $\{Z_{\myvec{a}_1}, \cdots,  Z_{\myvec{a}_l}\}$, there exists an efficient classical algorithm that returns the expectation values $\{\langle Z_{\myvec{a}_1}\rangle, \cdots,  \langle Z_{\myvec{a}_l}\rangle\}$ in  $\mathcal{O}{(mln)}$ runtime.
\end{proposition}
Notice that this algorithm returns the exact values of the expectation values rather than unbiased estimates as in Proposition~\ref{prop:expval_batch}.

\noindent \emph{Proof}: We start by deriving a general expression for expectation values $\langle Z_{\myvec{a}}\rangle$ of parameteized IQP circuits. Note that since $Z \exp(i\theta X) = \exp(-i\theta X) Z$ due to anticommutation of $Z,X$, by writing $U(\ttheta)=\prod_j \exp(i\theta_j X_{\myvec{g}_j})$ we have 
\begin{align}
    \langle Z_{\myvec{a}}\rangle &= \bra{0}U^\dagger(\ttheta)Z_{\myvec{a}}U(\ttheta)\ket{0} \\
    &=\bra{0}\prod_{j\vert \{X_{\myvec{g}_j},Z_{\myvec{a}}\}=0}\exp(-2i\theta_j X_{\myvec{g}_j})\ket{0} \\
    &=\bra{0}\prod_{j\vert \{X_{\myvec{g}_j},Z_{\myvec{a}}\}=0}(\cos(2\theta_j)\mathbb{I}+i\sin(-2\theta_j)X_{\myvec{g}_j})\ket{0}.
\end{align}
From this we can see that the only terms that survive when expanding the product are those that result in an identity operator. We therefore have 
\begin{align}\label{eq:expval_intefere}
    \langle Z_{\myvec{a}} \rangle = \prod_{j\vert \{X_{\myvec{g}_j},Z_{\myvec{a}}\}=0} \cos(2\theta_j) + \sum_{\omega \in \Omega}\prod_{j \notin \omega}\cos(2\theta_j)\prod_{j \in \omega}i\sin(-2\theta_j)
\end{align}
where $\Omega = \{\omega_1, \omega_2,\cdots\}$ is the collection of sets of indices $\omega\subseteq \{1,\cdots,m\}$ such that $\prod_{j\in \omega} X_{\myvec{g}_j} = \mathbb{I}$ holds for each $\omega\in\Omega$. We note that brute force calculation of this quantity is generally not possible since there may be exponentially many terms.

To derive a similar expression for stochastic bitflip circuits, we note that these circuits can be constructed as special cases of parameterized IQP circuits. Namely, given a parameterized IQP circuit, consider another parameterized IQP circuit with $n+m$ qubits, where $m$ is the total number of gates in the original circuit. For every gate with generator $X_{\myvec{g}_j}$ in the original, add a gate with generator $X_{\myvec{g}_j} X_{n+j}$ to this larger circuit, i.e.\ add an $X$ to each generator on a unique ancilla qubit. 

We now consider expectation values $Z_{\myvec{a}}$ acting on the first $n$ registers of this larger circuit. Note that the action of a gate with generator $X_{\myvec{g}_j} X_{n+j}$ on state $\ket{\xx}\ket{0}^{j}$ is to prepare the state
\begin{align}
    \cos(\theta_j)\ket{\xx}\ket{0}^{(j)} + i\sin(\theta_j)X_{\myvec{g}_j}\ket{\xx}\ket{1}^{(j)}.
\end{align}
As we do not measure the ancilla qubits, we may trace them out, so that the map is equivalent to 
\begin{align}
    \cos^2(\theta_j)\ket{\xx}\bra{\xx} + \sin^2(\theta_j)X_{\myvec{g}_j}\ket{\xx}\bra{\xx}X_{\myvec{g}_j} =  \cos^2(\theta_j)\ket{\xx}\bra{\xx} + \sin^2(\theta_j)\ket{\xx\oplus\myvec{g}_j}\bra{\xx\oplus\myvec{g}_j},
\end{align}
i.e. a stochastic bit flip of those bits acted on by $X_{\myvec{g}_j}$, as in \eqref{eq:bitflip}. This IQP circuit is therefore equivalent to a stochastic bitflip circuit when considering the first $n$ qubits only, and so we may use  \eqref{eq:expval_intefere} to obtain an expression for expectation values. Since each generator in the circuit contains an $X$ operator that acts on a unique ancilla qubit, one sees the $\Omega$ must be the empty set, and we therefore have that
\begin{align}\label{eq:expval_bitflip}
    \langle Z_{\myvec{a}} \rangle = \prod_{j\vert \{X_{\myvec{g}_j},Z_{\myvec{a}}\}=0} \cos(2\theta_j)
\end{align}
describes the expectation value structure of stochastic bitflip circuits. Interestingly, this makes for a clear exhibition of the role of interference in parameterized IQP circuits via the additional terms appearing in \eqref{eq:expval_intefere}. 

Due to the form of \eqref{eq:expval_bitflip} it follows that there is also an efficient classical algorithm to batch evaluate these expectation values. Using the same notation as Sec.~\ref{sec:batchexpval}, we can calculate \eqref{eq:expval_bitflip} by taking the element-wise cosine of the matrix $C\cdot \Theta$ and multiplying the values column-wise, resulting in a faster runtime of $\mathcal{O}{(mln)}$. $\blacksquare$

The package contains an option to use the stochastic bitflip model instead of the quantum model, as we will explain in Sec.~\ref{sec:overview}.  

\section{Package overview}\label{sec:overview}
We now give an overview of how to use the package with some simple example code. We note that while we generally use language from optimization in the following, we will sometimes adopt equivalent language from machine learning when the context requires. That is, we will sometimes use the term \emph{training} in place of \emph{optimization}, and \emph{loss function} in place of \emph{objective function}, and these terms should be understood to be equivalent.

\subsection{Creating and measuring a circuit}\label{sec:circuit}
Here we show how to create circuits, estimate their expectation values and probabilities, and sample from their output distributions. 

\begin{figure*}[t]
\begin{center}
\includegraphics[scale=.8]{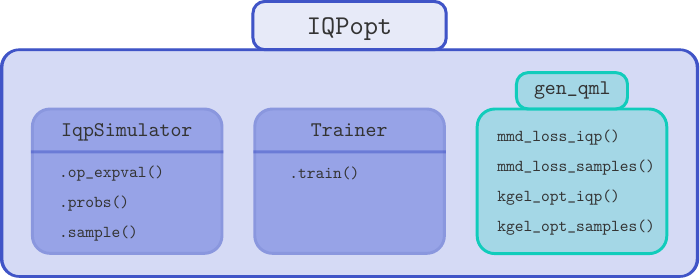}
\caption{Diagram of the structure of the package. While \python{IqpSimulator} and \python{Trainer} are classes, \python{gen\_qml} is a module that contains some functions useful for generative machine learning.}
\label{fig:diagram}
\end{center}
\end{figure*}

\subsubsection{Creating a circuit}

To define a parameterized IQP circuit we first need to specify the number of qubits and the gates that form the circuit. A set of gates is specified by a list.

\begin{center}
\python{gates = [gen1, gen2, gen3]}
\end{center}

that contains the generators of the gates. Generators are specified by their representation as binary vectors $\myvec{g}_j$ as in \eqref{eqn:expvalsample}. For example, \python{gen1 = [0,1]} corresponds to a gate with generator $X\otimes X$ acting on the first two qubits of the circuit. For a three qubit system the corresponding parameterized gate is therefore $\exp(i\theta_1 X\otimes X\otimes \mathbb{I})$. 

The following example prepares a three qubit circuit with all single and two qubit gates via the function \python{local\_gates}, i.e.

\begin{center}
\python{gates = [[0], [1], [2], [0,1], [0,2], [1,2]]}. 
\end{center}

\vspace{10pt}
\centerline{\includegraphics[width=1.0\textwidth]{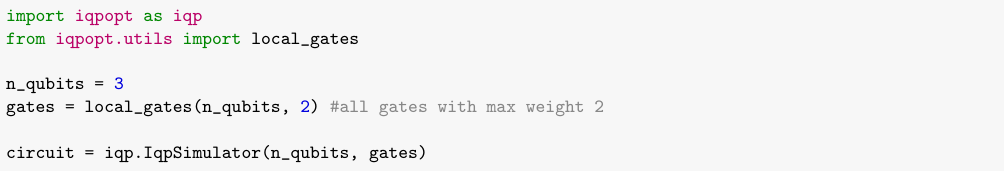}}

To use sparse operators to control memory (Sec.\ \ref{sec:memory}), the optional argument \python{sparse=True} can be passed to \python{IqpSimulator}, and to define a stochastic bitflip circuit (Sec.~\ref{sec:bitflip}) we set \python{bitflip=True}. 

In the above, each gate has a unique optimization parameter. It is also possible to assign the same parameter to more than one generator. To do this we represent a gate by a tuple of generators \python{(gen1,gen2,...)}. As a concrete example, for a three qubit circuit the gates

\begin{center}
\python{gates = [([0,2], [0,1,2]), ([0],)]}.
\end{center}

correspond to the parameterized unitaries
\begin{align}
    \exp(i\theta_1 (X \otimes \mathbb{I} \otimes X+X \otimes X \otimes X)) \quad\text{and}\quad\exp(i\theta_2 X \otimes \mathbb{I} \otimes \mathbb{I}).
\end{align}
It is also possible to specify a number of initial gates that have fixed parameters that are not optimized. This is done by the optional arguments \python{init\_gates} and \python{init\_coefs}, which specify a list of gates and the values of their fixed parameters.

\subsubsection{Estimating expectation values}\label{sec:expvals}
The class method \python{IqpSimulator.op\_expval()} is used to provide estimates of expectation values via the equation \eqref{eqn:expvalsample}.  The function requires a parameter array containing the values $\theta_j$, an operator $\langle Z_{\myvec{a}} \rangle$ specified by its binary representation $\myvec{a}$, the number of samples $s$ that controls the precision (the more the better), and a JAX pseudo random number generator key to seed the randomness of the sampling. Below is a simple example implementing this for the three qubit circuit defined above, which returns the expectation value estimate as well as its standard error.

\vspace{10pt}
\centerline{\includegraphics[width=1.0\textwidth]{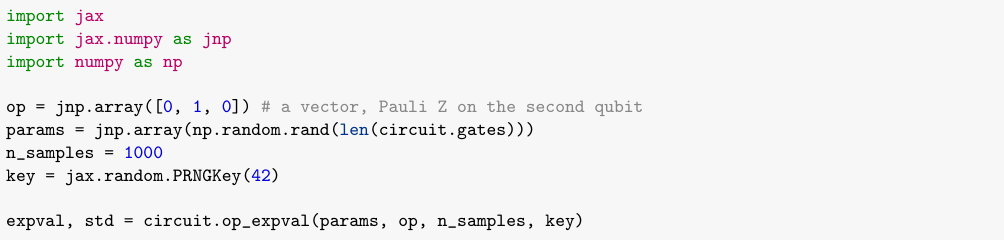}}

The function also allows for fast batch evaluation of expectation values following the method in Sec.~\ref{sec:batchexpval}. If we specify a batch of $Z_{\myvec{a}}$ operators by an array

\vspace{10pt}
\centerline{\includegraphics[width=1.0\textwidth]{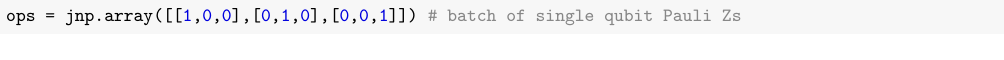}}

\vspace{-15pt}

we can batch evaluate the expectation values and standard deviations in parallel with the same syntax

\vspace{10pt}
\centerline{\includegraphics[width=1.0\textwidth]{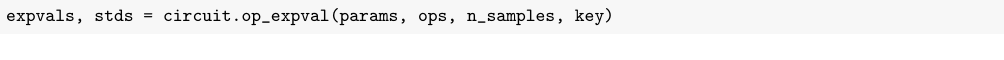}}
\vspace{-15pt}

As described in Sec.~\ref{sec:batchexpval}, this uses the same randomness for each operator in the batch, which may result in correlations between estimates. If independent randomness is needed, it can be specified with the optional argument \python{indep\_estimates=True} at the cost of longer runtime. Memory usage can be controlled by the optional arguments \python{max\_batch\_ops} and \python{max\_batch\_samples}, which correspond to the batching of operators and bit string samples described in Sec.~\ref{sec:memory}.

\subsubsection{Sampling and probabilities}

\python{IqpSimulator} has two methods that make use of state vector simulation in the quantum computing library PennyLane \cite{pennylane}. Since these rely on state vector simulation, they do not scale efficiently with \python{n\_qubits}.

\begin{itemize}
    \item \python{.sample()} samples bit strings from the circuit output, and takes as arguments the circuit parameters \python{params} and the number of required shots \python{shots}. For circuits with \python{bitflip=True}, this method does scale efficiently with \python{n\_qubits} since sampling can be performed classically. 

    \item \python{.probs()} returns an array containing the probabilities of sampling each possible bitstring, and takes \python{params} as argument. 
\end{itemize}

If more complex tasks in PennyLane are required, the circuit expressed as a function in PennyLane can be accessed via the class attribute \python{.iqp\_circuit}.

\subsection{Optimizing a circuit}\label{sec:circuitopt}

Circuits can be optimized via a separate \python{Trainer} class. To instantiate a trainer object we first define a loss function (also called an objective function), an optimizer and an initial stepsize for the gradient descent. Continuing our example from before, below we define a simple loss function that is a sum of expectation values returned by \python{op\_expval}.

\vspace{10pt}
\centerline{\includegraphics[width=1.0\textwidth]{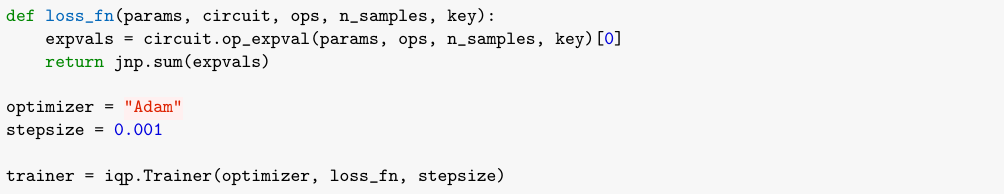}}

Any differentiable loss function expressible in JAX can be defined, but must have a first argument \python{params} that corresponds to the optimization parameters of the circuit. Two other optimizers (\python{"GradientDescent"} and \python{"BFGS"}) are also available. If the optional argument \python{opt\_jit=True} is passed, the entire optimization is just-in-time complied, however requires that all the arguments of the loss function are traceable by JAX.

To minimize the loss function, we call the method \python{.train()}, which requires the number of iterations \python{n\_iters} and the initial arguments of the loss function \python{loss\_kwargs} given as a dictionary object. This dictionary must contain a key \python{params} whose corresponding value specifies the initial parameters. 

\vspace{10pt}
\centerline{\includegraphics[width=1.0\textwidth]{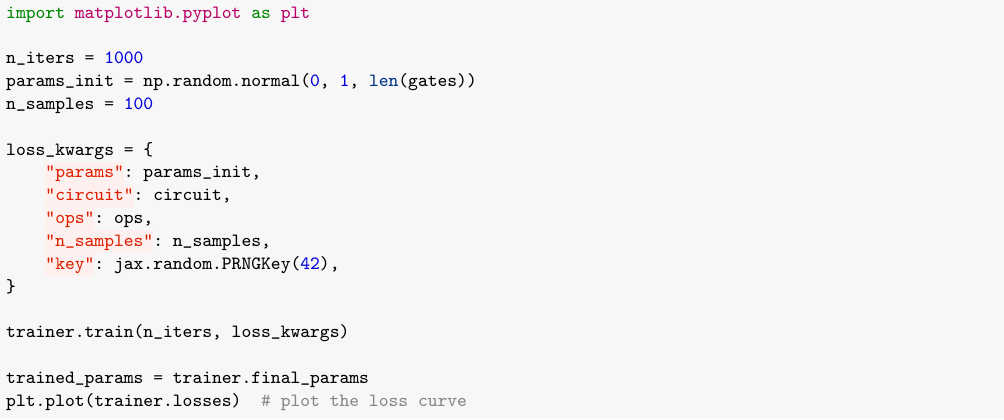}}

There are a number of additional optional arguments that can be passed to \python{.train()}:

\begin{itemize}
    \item \python{convergence\_interval} is the number of steps over which to decide convergence. If the loss does not decrease, or increases over this scale, the training is stopped. Defaults to \python{None} (no convergence criterion is used).
    \item \python{val\_kwargs} must be a dictionary with the same keys as \python{loss\_kwargs} except the key \python{params}. The value of the loss with these values and the current value of \python{params} is monitored to decide convergence if \python{convergence\_interval=True}. This allows one to construct a validation loss, as is common in machine learning. If \python{None}, the training loss is used to decide convergence. Defaults to \python{None}.
    \item \python{monitor\_interval} is an integer which if specified, saves the variable \python{params} in a class attribute called \python{.params\_hist} every \python{monitor\_interval} steps. Defaults to \python{None}, meaning there is no monitoring.
    \item \python{turbo} is an integer which if specified jit compiles the parameter update step and uses jax.lax to compile every \python{turbo} number of iterations for faster training. This is only compatible with circuits for which \python{sparse=False}. 
    \item \python{random\_state} is an integer that is used as a seed for random functions used while training. If \python{loss\_kwargs} contains a key called \python{key}, this is used instead. 
\end{itemize}

\subsection{Generative machine learning tools}\label{sec:gen}
Generative machine learning is a subfield of machine learning that uses data to model and sample from high dimensional probability distributions. Since quantum circuits define probability distributions at their output, we can view a parameterized IQP circuit as a generative quantum machine learning model that generates samples of binary vectors according to the distribution
\begin{align}
q_{\ttheta}(\xx) \equiv q(\xx\vert\ttheta)=\vert\bra{\xx}U(\ttheta)\ket{0}\vert^2.
\end{align}
The package contains a module \python{gen\_qml} with functionality to train and evaluate generative models expressed as \python{IqpSimulator} circuits. The technical proofs which accompany this code can be found in an upcoming publication \cite{upcoming} that leverages the package to investigate generative machine learning with parameterized IQP circuits. 

\subsubsection{Training via the maximum mean discrepancy loss}\label{sec:mmd}

The Maximum Mean Discrepancy (MMD) \cite{JMLR:v13:gretton12a} is an integral probability metric that measures the similarity between two probability distributions, and can serve as a loss function to train generative models. For a ground truth distribution $p$ and a generative model distribution $q_{\ttheta}$, the loss we consider is the square of the MMD, given by
\begin{align}
    \text{MMD}^2(\ttheta)
    = \, & \bbE_{\xx,\yy\sim q_{\ttheta} }[k(\xx,\yy)] - 2  \bbE_{\xx \sim q_{\ttheta},\yy\sim p }[k(\xx,\yy)] + \bbE_{\xx,\yy\sim p }[k(\xx,\yy)] \,,
    \label{eq:mmd_loss}
\end{align}
where $k$ is the Gaussian kernel $k(\xx, \yy)=\exp(\vert\vert x - y\vert\vert^2/2\sigma^2)$, and $\sigma$ is called the bandwidth. The package contains two functions that can be used to obtain estimates of MMD$^2$: \python{mmd\_loss\_samples} and \python{mmd\_loss\_iqp}.

When $q_{\ttheta}$ is a generative model from which we can obtain samples, the function \python{mmd\_loss\_samples} returns an unbiased estimate of MMD$^2$ following the method in \cite{JMLR:v13:gretton12a} (lemma 6), given a set of samples from $p$ and from $q_{\ttheta}$. In the following example we estimate the MMD$^2$ of two similar distributions over low weight bitstrings. We use the median heuristic \cite{medianheuristic} to set the bandwidth parameter. 

\vspace{10pt}
\centerline{\includegraphics[width=1.0\textwidth]{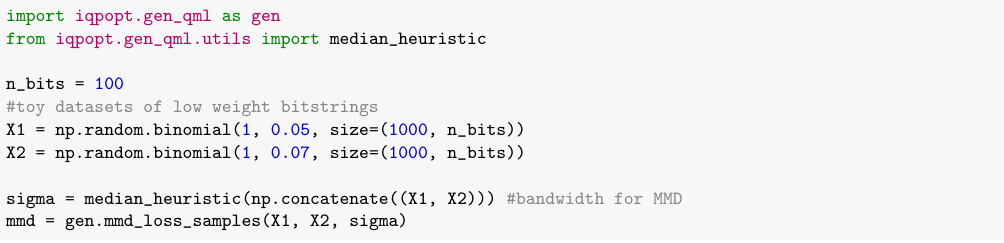}}

When $q_{\ttheta}$ is an \python{IqpSimulator} object we can use \python{mmd\_loss\_iqp} to obtain unbiased estimates of the MMD$^2$. For the case of a Gaussian kernel with bandwidth $\sigma$, it is shown in \cite{rudolph2024trainability} (see supplementary note 4) that the loss can be written as an expectation with respect to a known distribution $\mathcal{P}_{\sigma}(\myvec{a})$:
\begin{align}\label{eqn:mmd_estimate}
    \text{MMD}^2(\ttheta)
    = \mathbb{E}_{\myvec{a} \sim\ \mathcal{P}_{\sigma}(\myvec{a})}\Big[ \langle Z_{\myvec{a}} \rangle_p \langle Z_{\myvec{a}} \rangle_{p} - 2\langle Z_{\myvec{a}} \rangle_p \langle Z_{\myvec{a}} \rangle_{q_{\ttheta}} +\langle Z_{\myvec{a}} \rangle_{q_{\ttheta}} \langle Z_{\myvec{a}} \rangle_{q_{\ttheta}}\Big]
\end{align}

where $\langle Z_{\myvec{a}} \rangle_{p}=\mathbb{E}_{\xx\sim p}[(-1)^{\myvec{a}\cdot\xx}]$ is the expectation value of $Z_{\myvec{a}}$ on the ground truth distribution. Since this expression involves expectation values only, it is possible use \python{op\_expval} to construct an unbiased estimator of MMD$^2$ \cite{upcoming} which is implemented via the function \python{mmd\_loss\_iqp} as follows.

\vspace{10pt}
\centerline{\includegraphics[width=1.0\textwidth]{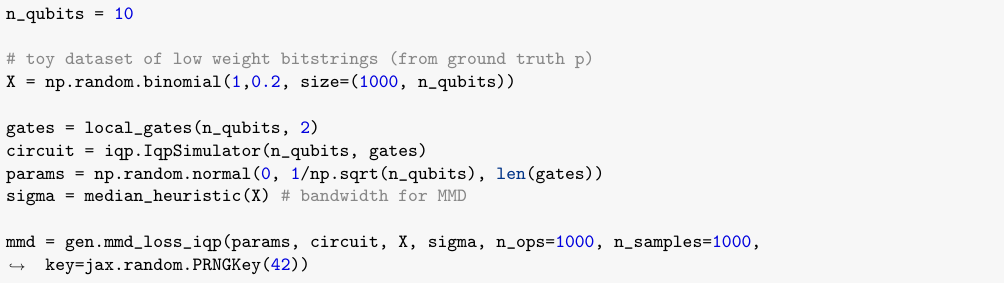}}

The variable \python{n\_ops} controls the number of operators used to estimate the MMD$^2$ by converting \eqref{eqn:mmd_estimate} to an empirical mean over a sample of operators from $\mathcal{P}_{\sigma}(\myvec{a})$, and therefore affects the precision of the estimate. The argument \python{n\_samples} also affects precision since it controls the precision of the estimates of each expectation value via \python{op\_expval}.

This function can be used with a \python{Trainer} object to train a quantum generative model given as a parameterized IQP circuit in a similar fashion to the example of Sec.~\ref{sec:circuitopt}.

\vspace{10pt}
\centerline{\includegraphics[width=1.0\textwidth]{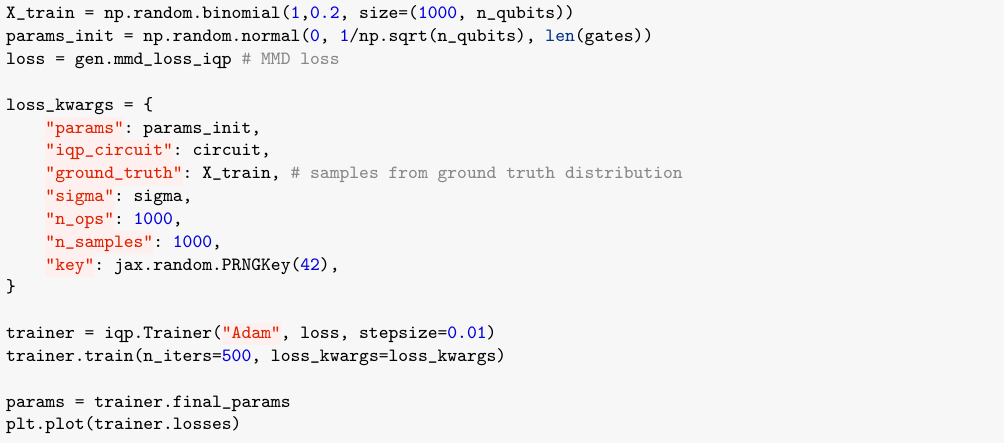}}

As before, memory can be controlled by specifying keywords \python{max\_batch\_ops} and \python{max\_batch\_samples} which are passed to \python{op\_expval}, and \python{indep\_estimates=True} can be used to force independent randomness in the estimation of expectation values (at the cost of larger runtime).\footnote{Note that  even for \python{indep\_estimates=False} the function has been designed to return an unbiased estimator, so it is unclear if this options is useful in practice.}

There are several optional arguments that can be defined when calling \python{mmd\_loss\_iqp}:

\begin{itemize}
    \item \python{wires} is a list specifying the qubits to sample from in the IQP circuit, treating the rest of the qubits as ancillas. For example, \python{wires=[0,1]} samples from the first two qubits only. By default, all qubits are treated as wires.
    \item \python{jit} is a boolean that controls whether to just-in-time compile the loss or not. Works only for circuits with \python{sparse=False}. Defaults to \python{True}.
    \item \python{sqrt\_loss} is a boolean that controls whether to use the square root of the MMD$^2$ loss. Note that estimates will no longer be unbiased. Defaults to \python{False}.
\end{itemize}

\subsubsection{Kernel Generalized Empirical Likelihood}
The package also contains functions to estimate the Kernel Generalized Empirical Likelihood \cite{ravuri2023understandingdeepgenerativemodels}: a recently introduced tool that is particularly useful for diagnosing failures of learning such a mode dropping and mode imbalance.  

The precise metric we focus on is the KGEL test from \cite{ravuri2023understandingdeepgenerativemodels}, where we have taken the feature map to be the trivial identity map\footnote{specifying a non-trivial feature map can generally be useful for high dimensional data, however it is unclear if the techniques here can be adapted to work in this case.}:
\begin{equation}
\text{KGEL} = 
\begin{gathered}
\min_{\{\boldsymbol{\pi} | \sum_i \pi_i=1, \pi_i \geq 0\}} D_{KL}(P_{\boldsymbol{\pi}}\vert\vert\hat{P}) \quad \mbox{subject to} \quad
\sum_{i=1}^N \pi_i 
{\scriptscriptstyle \begin{bmatrix}
k(\xx_i, t_1)\\
\vdots \\
k(\xx_i, t_W)
\end{bmatrix}}
=
\mathbb{E}_{\myvec{y}\sim q_{\ttheta}} \begin{bmatrix}
k(\yy, t_1)\\
\vdots \\
k(\yy, t_W)
\end{bmatrix}.%
\end{gathered}
\end{equation}

Here the $\xx_i$ are $N$ samples drawn from the ground truth distribution, $\hat{P}$ is the empirical distribution $\hat{P}(\xx) = \sum_i N^{-1}\mathbb{I}_{[\xx_i=\xx]}$, the points $\boldsymbol{t}_i$ are ``witness points'' typically sampled from the ground truth, and we take the kernel to be the Gaussian kernel as before. Note that in the limit of large $N$, if $p=q_{\ttheta}$, the constraint becomes an equality if $P_{\boldsymbol{\pi}}=\hat{P}$, minimizing the KL divergence. If a particular mode is dropped by the generative model, then less probability will generally need to be assigned to those $\pi_i$ that belong to that mode, which can be detected by inspecting the optimal solution $\boldsymbol{\pi}^*$ (see \cite{ravuri2023understandingdeepgenerativemodels} for more details). In this form the KGEL is also a convex optimization problem, which we solve with the pacakge \texttt{cvxpy} \cite{cvxpy1,cvxpy1}. The module contains two functions for estimating the KGEL: \python{kgel\_opt\_samples} and \python{kgel\_opt\_mmd}. 

As with \python{mmd\_loss\_samples}, the function \python{kgel\_opt\_samples} can be used only if samples are available from the generative model. An example usage is as follows.  

\vspace{10pt}
\centerline{\includegraphics[width=1.0\textwidth]{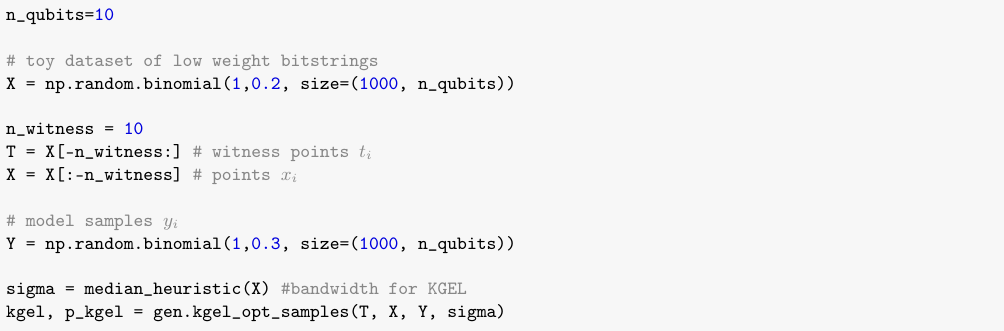}}

The value of the KGEL, as well as the optimal distribution $\boldsymbol{\pi}^*$ are returned. 

To evaluate the KGEL for a model given by a \python{IqpSimulator} object, we need to use the function \python{kgel\_opt\_iqp}. Similarly to the MMD$^2$, this leverages \python{op\_expval} to provide an estimate and can be used as follows (see \cite{upcoming} for a proof).

\vspace{10pt}
\centerline{\includegraphics[width=1.0\textwidth]{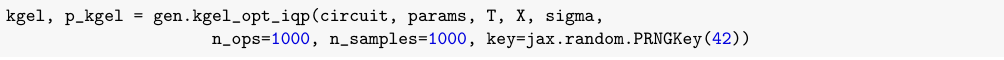}}

As usual, higher precision can be requested by increasing \python{n\_ops} and \python{n\_samples}, and the options \python{wires}, \python{indep\_estimates}, \python{max\_batch\_samples} and \python{max\_batch\_ops} are available with analogous behavior to before. We remark that this function does not return an unbiased estimate of the KGEL, and can require considerable precision to arrive at stable KGEL values. For large problems \python{n\_ops} and \python{n\_samples} may therefore need to be set to significantly larger values.

\section{Benchmarks }\label{sec:benchmarks}
In this section we investigate the limits of the package via a number of benchmark tests. These tests were performed on a NVIDIA Grace Hopper GH200 chip, which features an NVIDIA Tensor Core GPU coupled via a high bandwidth link to a 72 core CPU, providing a total combined memory of 624 GB. We investigate three implementations:
\begin{enumerate}
    \item A CPU-only implementation, using the option \python{Sparse=False}.
    \item A CPU-only implementation, using the option \text{Sparse=True}.
    \item An implementation with full CPU and GPU access, using the option \python{Sparse=False}. GPU usage is controlled via the default behavior of JAX without any GPU-specific code optimization. 
\end{enumerate}
For each benchmark, we consider circuits with either 1000 (1k), 100000 (100k) or 1000000 (1m) gates. Gate generators are chosen randomly such that, for each generator, the probability that the generator has an $X$ operator on a given qubit is $3/n$. This therefore results in a set of gates with an average Pauli weight of 3, whose sparseness may be exploited by sparse matrix operations in the second implementation. We note this choice is somewhat natural since quantum circuit gates tend to act on few qubits.

\begin{figure*}
  \centering
  \includegraphics{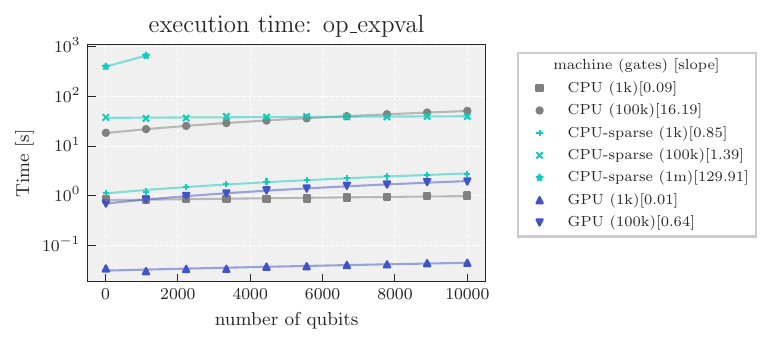}
  \caption{Runtimes for the batch estimation of 10000 randomly chosen observables, for different implementations and circuit gate counts. The solid lines are linear fits of time as a function of qubit number, with the corresponding slope given in the legend. The gates and observables are chosen randomly such that their average Pauli weight is three, which allows the sparse implementation to scale to 1 million gates at the cost of longer runtimes. The first points in the plot corresponds to a circuit of 100 qubits. \label{fig:bench_opexpval}}
\end{figure*}

\begin{figure*}
\centering
  \centering
  \includegraphics[width=1.0\textwidth]{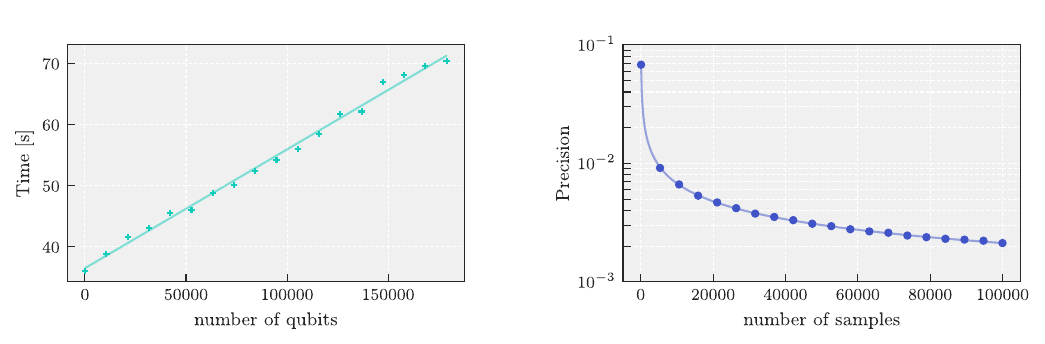}
  \captionof{figure}{(left) Execution time for the benchmark of Sec.~\ref{sec:bench_opexpval} (for 100k gates) for the sparse implementation. The runtime is well described by a linear fit even for circuits with up to 200k qubits. (right) Scaling of the average precision of the estimates returned by \python{op\_expval} as a function of the argument \python{n\_samples} for the operators used in benchmark tests (the precision is largely independent of qubit number or gate count). The line is a linear fit using the variable $1/\sqrt{\python{n\_samples}}$. Note that the same scaling occurs when estimating expectation values from samples of the quantum circuit.}
  \label{fig:bench_precision}
\end{figure*}

\subsection{Expectation values}\label{sec:bench_opexpval}
We first investigate the execution of \python{op\_expval} by performing a batch evaluation of 10000 expectation values (see Fig.~\ref{fig:bench_opexpval}). The operators $Z_{\myvec{a}}$ in this batch are chosen via the same procedure as the gate generators, so that their average Pauli weight is again three. We set \python{n\_samples=10000}, which amounts to an additive precision of around 0.007 for each estimate; see Fig.~\ref{fig:bench_precision} for a plot of how precision scales with \python{n\_samples}. 

As we can see from Fig.~\ref{fig:bench_opexpval}, the calculation benefits significantly from GPU acceleration, which results in a factor 25 speedup for circuits with 1k and 100k gates. Both the GPU and CPU implementations failed to execute for the case of 1m gates due to insufficient memory; the CPU-sparse implementation was able overcome this limit, however also exhausted the memory at 2000 gates. 

The scaling of runtime is well described by a linear fit in all cases. For the CPU and GPU implementations this is expected from the analysis of Sec.~\ref{sec:batchexpval} and Sec.~\ref{sec:memory}. For the sparse implementation, the quadratic dependence on the number of qubits is not apparent at this scale. In Fig.~\ref{fig:bench_precision} (left) we extend the benchmark for the sparse implementation (for the case of 100k gates) up to 200k qubits. Interestingly, the runtime is still well described by a linear fit even at this scale, and so the sparse implementation appears to scale well even for very large circuits. The sparse implementation also becomes faster than the standard CPU implementation for circuits with more that 7000 qubits. This can be seen more clearly in Fig.~\ref{fig:intro}, which features a linear time axis. 

We remark that proper exploitation of sparsity in this problem would likely lead to significantly faster runtimes. In the current code, the calculation of the matrix $E$ in \eqref{eqn:Emat} is performed with standard matrix multiplication in JAX. This is necessary in order to be compatible with the automatic differentiation pipeline, since gradients flow through this computational node (see Fig.~\ref{fig:comp_graph}). However, one could evaluate and code the gradient of \eqref{eqn:expvalsample} by hand, avoiding the need for automatic differentiation, and therefore permitting the use of sparse operations when differentiating through $E=C\cdot\Theta\cdot B^T$. Since the matrix $C$ will be sparse for sparse gates and operators, savings can therefore be made here. Combining this technique with dedicated high-performance-computing software for sparse matrix operations (such as cuSparse \cite{cuSPARSE}) may be crucial to scale optimizations to millions of qubits and gates.

\begin{figure*}
    \includegraphics[width=\textwidth]{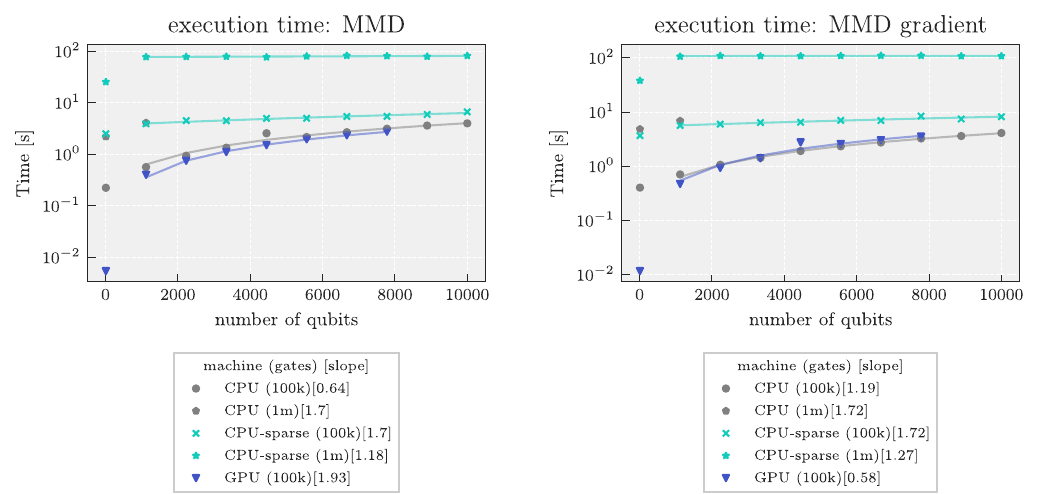}
    \caption{Runtime to estimate the squared maximum mean discrepancy of Eq.~\eqref{eq:mmd_loss} and its gradient via the function \python{mmd\_loss\_iqp} for circuits of varying sizes and gate counts, and setting \python{n\_samples = 1000} and \python{n\_ops = 1000}. Runtimes obey an approximately linear dependence of qubit number for circuits larger than 1000 qubits, given by the solid lines. The first points in the plot corresponds to a circuit of 100 qubits. \label{fig:bench_mmd}}
\end{figure*}

\subsection{MMD loss}
Here we probe the execution time of the function \python{mmd\_loss\_iqp} described in Sec.~\ref{sec:mmd}. We chose \python{n\_ops=1000} and \python{n\_samples=1000}, opting for a smaller value of \python{n\_samples} than previously since stochastic gradients in machine learning do not typically have to be very precise. In our own experiments, we have found this choice to be suitable to obtain good results for large-scale machine learning problems. 

In Fig.~\ref{fig:bench_mmd} we show the runtimes to evaluate the MMD$^2$ and its gradient for circuits with 100k or 1m gates. After 1000 qubits the scaling behavior again approximately obeys a linear relationship, since \python{op\_expval} is the main computational bottleneck in this calculation. We see that lower values of \python{n\_ops} and \python{n\_samples} compared to the previous benchmark free enough memory to enable million-gate circuits for both CPU implementations, with the sparse implementation now scaling to 10000 qubits and 1m gates. We also see that the execution time of the gradient is similar to that of the MMD$^2$, as can be expected from automatic differentiation. 

Interestingly, the use of GPU acceleration does not lead to significant advantages at this scale, and exhausts the memory before the CPU implementation, however we expect this situation could be improved with code optimization targeted at GPU hardware. We also note that for larger values of \python{n\_ops} and \python{n\_samples} we would expect better performance relative to CPU implementations from the results of the previous benchmark and the use of \python{op\_expval} within this function. 

\section{Conclusion}\label{sec:conlcusion{}}
In this work we have shown that optimization of large quantum circuits is not only possible, but is achievable today with only classical resources. This raises the prospect of investigating the power of large-scale quantum computation via numerical studies carried out in IQPopt, which may be able to provide insights beyond the limited apparatus of pen-and-paper analysis. 

Although variational in nature, our approach is compatible with fault tolerant quantum computing since rotation gates can be efficiently compiled to universal gate sets via the Solovay–Kitaev theorem \cite{kitaev1997quantum}. We therefore envision that IQPopt eventually become a useful component in fault tolerantly compiled algorithms. Ultimately, we believe the success of this approach will depend on identifying relevant problems that can be cast efficiently as IQPopt problem instances, as is already the case for certain quantum machine learning and quantum optimization tasks. Further research in this direction, as well as improvements that extend the sets of states and objective functions to which the package applies, will be the subject of future research. 

\section{Acknowledgments}
We thank the authors of the package \href{https://github.com/johannesjmeyer/rsmf}{rsmf} for contributing to the nice figures of this paper. JB acknowledges useful comments and  
conversations with David Wakeham, Maria Schuld, Shahnawaz Ahmed, Alexia Salavrakos and Patrick Huembeli. JB thanks Lee O'Riordan and Sanchit Bapat for help with HPC resources. This work has been supported by the Government of Spain (Severo Ochoa CEX2019-000910-S, FUNQIP and European Union NextGenerationEU PRTR-C17.I1), Fundació Cellex, Fundació Mir-Puig, Generalitat de Catalunya (CERCA program) and European Union (PASQuanS2.1, 101113690). EA is a fellow of Eurecat's \textit{Vicente López} PhD grant program and is grateful to Antonio Acín's and Adan Garriga's support with this work and his PhD.

\printbibliography

\onecolumn
\appendix

\section{Runtime analysis for the sparse implementation}\label{app:sparse}
The SciPy package implements the SMPP algorithm for fast matrix multiplication found in \cite{smpp}. For the matrix multiplication $M\cdot N$, where $M$ is of dimension $m\times k$ and $N$ of dimension $k\times n$ this algorithm has runtime complexity
\begin{align}
    \mathcal{O}(m[\max\left(\text{nnz}_{\text{row}}(M), \text{nnz}_{\text{col}}(N)\right)^2+ \max(m,n))
\end{align}
where $\text{nnz}_{\text{row}}(M)$ and $\text{nnz}_{\text{col}}(N)$ are the maximum number of non-zero entries in any row of M and column of $N$. 

Using this, we can now consider each of the matrix multiplications needed to compute the batch of expectation values. The calculation of $A\cdot G^T$ will require 
\begin{align}
    \mathcal{O}(l\max\left(\text{nnz}_{\text{row}}(A), \text{nnz}_{\text{row}}(G)\right)^2+\max(l,m))
\end{align}
and the calculation of $Z\cdot G^T$ will require 
\begin{align}
    \mathcal{O}(s n^2+\max(s,m))
\end{align}
time. In the above we have assumed $Z$ to be dense, and hence the maximum number of non-zero entries in a row scales as $n$. 

Since we require gradients in JAX, the remaining operations must be computed with standard matrix multiplication. We thus convert $Z\cdot G^T$ to dense matrix format in $\mathcal{O}(m\cdot s)$ runtime (note $Z\cdot G^T$ will not be sparse since $Z$ is dense and random), and create the dense matrix $C=1-(-1)^{A\cdot G^T}$ in runtime $\mathcal{O}(\text{nnz}(A\cdot G^T))$, where $\text{nnz}$ denotes the total number of non-zero entries in the matrix. Finally we use dense matrix multiplication to compute $E=C\cdot\Theta\cdot B^T$ in runtime $\mathcal{O}(sml)$. This results in a total runtime
\begin{align}
    \mathcal{O}(l\max\left(\text{nnz}_{\text{row}}(A), \text{nnz}_{\text{row}}(G)\right)^2 + s n^2 + sml).
\end{align}
Since we need to convert matrices to dense format for JAX compatibility, the memory cost will not be dramatically different to the non-sparse implementation. The main memory saving with the sparse implementation is the fact that the matrix $G$ is not stored as a dense matrix, and therefore has memory cost $\mathcal{O}(nnz(G))$ rather than $mn$. For circuits with large numbers of gates and qubit relative to $s$ and $l$, this option can therefore be beneficial.

\end{document}